\RequirePackage{fix-cm}
\documentclass[smallextended]{svjour3}       
\smartqed  
\usepackage{graphicx}
\begin{document}

\title{Dimensional effects in Efimov physics 
}


\author{M. T. Yamashita}


\institute{M.T. Yamashita \at
           Instituto de F\'\i sica Te\'orica, Universidade Estadual Paulista, UNESP, Rua Dr. Bento Teobaldo Ferraz, 271 - Bloco II, 01140-070 S\~ao Paulo, SP, Brazil \\
           Tel.: +55-11-33937830\\
              \email{marcelo.yamashita@unesp.br}          
}

\date{Received: date / Accepted: date}

\maketitle

\begin{abstract}
Efimov physics is drastically affected by the change of spatial dimensions. Efimov states occur in a tridimensional ($3$D) environment, but disappear in two ($2$D) and one ($1$D) dimensions. In this paper, dedicated to the memory of Prof. Faddeev, we will review some recent theoretical advances related to the effect of dimensionality in the Efimov phenomenon considering three-boson systems interacting by a zero-range potential. We will start with a very ideal case with no physical scales \cite{derickpra}, passing to a system with finite energies in the Born-Oppenheimer (BO) approximation \cite{derickjpb} and finishing with a general system \cite{john}. The physical reason for the appearance of the Efimov effect is given essentially by two reasons which can be revealed by the BO approximation - the form of the effective potential is proportional to $1/R^2$ ($R$ is the relative distance between the heavy particles) and its strength is smaller than the critical value given by $-(D-2)^2/4$, where $D$ is the effective dimension.
\end{abstract}

\section{Introduction}
\label{intro}
Efimov physics is an important branch of the Few-Body community with articles in several research areas, some of them very unusual \cite{dna}. The so-called Efimov physics received a special attention since Efimov states \cite{efimov} have been indirectly identified inside ultracold atomic traps \cite{grimm}. The possibility to tune the two-body scattering lengths by using the Feshbach resonance technique \cite{pethickbook} allows to place the three-body bound states exactly at the transition to the continuum. In this position, the trimers may recombine into deeply bound dimers plus one atom with large enough kinetic energies to allow them to escape from the trap \cite{braaten}. The peak corresponding to the loss of atoms is then associated with an Efimov state. This first detection, however, was related to an artificial production of molecules at the continuum threshold.

Several theoretical papers predicted the existence of genuine Efimov states appearing in nature \cite{kolganova}. The most promising candidate was detected in 2015 by the group headed by D\"orner \cite{dornertrimer}. In this article, the energy of the ground and first excited states of a trimer of $^4$He was measured directly. The ingenious technique consists firstly to prepare a molecular beam under an expansion of a gaseous He. By matter-wave diffraction it is possible to separate monomers, dimers, trimers, etc. Then, an ultrashort laser field ionize the atoms of the trimer causing a Coulomb explosion. All momenta of the ionized atoms can be measured and the geometry immediately before the ionization process can be reconstructed using Newton's equations. 

A direct measurement of the Efimov trimer energies and also the helium-dimer wave function \cite{dornerdimer} shows that a new aspect in few-body physics would be interesting to be studied in experiments: how the effective spatial dimension in atomic traps influence the observables related to Efimov physics. The question of dimensionality is frequently explored in the context of many atoms. Two or one dimensions can be produced by changing asymmetrically the magnetic and electric fields, then the 3D atomic cloud obtains a pancake or cigar shape \cite{petrov}. However, in Efimov physics the study of intermediate dimensions is still very incipient \cite{nielsen,tan,levinsen,christensen,naidon}.

It is well established that Efimov effect does not happen in 2D\footnote{Considering higher partial waves or more particle interactions, similar effects, with other scaling laws may, appear in dimensions different from 3D \cite{nishida}.}. The Efimov effect is the appearance of an infinite number of three-body bound states when the energy of at least two of the three two-body subsystems tend to zero. Nielsen and collaborators showed in Ref. \cite{nielsen} that the effect does not cease to exist exactly at 2D, but there is an interval in which Efimov effect may exist. For three equal mass bosons they found that this interval is given by $2.3<D<3.8$.  

In this article, prepared for this Faddeev special volume, we merged our recent results involving the question of dimensionality in few-body systems. In the following section, we will start with a very ideal situation where we consider only the region of large momenta \cite{derickpra}. This very simple procedure, based on the ideas of Danilov's article \cite{danilov}, extends the result of Ref. \cite{nielsen} to a mass imbalanced system. The situation is analogous of considering an infinite high excited Efimov state with all energies set equal zero. Then, there are no scales and the dimensions where the Efimov effect exists can be easily extracted. 

In Section 3 we show the physical reason for the disappearance of Efimov states \cite{derickjpb}. We use the BO approximation to generalize the result from Fonseca, Redish and Shanley \cite{fonseca} and obtain the effective potential in $D$ dimensions for  $AAB$ systems, which is responsible for the fall to the center phenomenon that generates the infinite number of three-body bound states. Finally, we consider a generic $AAB$ system with finite energies \cite{john} and show the energy spectrum for a continuous transitions from $3$D-$2$D-$1$D. 

\section{The ideal case: no physical scales}

Throughout this paper we will consider a three-body system, $AAB$, formed by two identical bosons, $A$, and a different particle, $B$, interacting by a pairwise Dirac-delta potential. The Efimov spectrum can be obtained in 3D by solving the Skorniakov and Ter-Martirosyan (STM) \cite{stm} coupled integral equations for the bound state of a system of three bosons \cite{raquelpra}. Here, we are replacing the tridimensional integrals, present in the STM equations, generalizing them to an arbitrary dimension $D$. Thus, the generalized STM coupled integral equations written in terms of the spectator functions $\chi_{A}$ and $\chi_{B}$ in units of $\hbar=m_A=1$ reads (part of the content of this section was originally published in Ref. \cite{derickpra})
\begin{small}
\begin{eqnarray}
\label{chia}
\nonumber
\chi_{A}(q) &=& \tau_{AB}\left(E_3 -\frac{{\cal A}+2}{2({\cal A}+1)}q^2\right) \\
&\times&\int d^{D}k\left(\frac{\chi_{B}(k)}{E_3 - q^2 - \frac{{\cal A}+1}{2{\cal A}}k^2 
- \vec{k}\cdot\vec{q}}+
\frac{\chi_{A}(k)}{E_3  - \frac{{\cal A}+1}{2{\cal A}}(k^2+q^2) 
- \frac{1}{{\cal A}}\vec{k}\cdot\vec{q}} \right) ,
\\ 
\label{chib}
\chi_{B}(q) &=& 2 \, \tau_{AA}\left(E_3 - \frac{{\cal A}+2}{4{\cal A}}q^2\right)  
\int d^{D}k \, \frac{\chi_{A}(k)}{E_3 - \frac{{\cal A}+1}{2{\cal A}}q^2 -k^2- 
\vec{k}\cdot\vec{q}}, 
\end{eqnarray}
\end{small}
where $E_3$ is the energy of the three-body system. The relative Jacobi momenta $\vec{q}$ and $\vec{k}$ 
are defined such that their origin is the center-of-mass of a given pair and point towards the remaining particle. The two-body transition amplitudes $\tau_{AB}$ and $\tau_{AA}$ are given by
\begin{small}
\begin{eqnarray}
\label{tauab}
\nonumber
&&\tau^{-1}_{AB}\left(E_3 -\frac{{\cal A}+2}{2({\cal A}+1)}q^2\right) = 
\int d^D k \left(\frac{1}{- |E_2^{AB}|- \frac{{{\cal A}}+1}{2{{\cal A}}}k^2} -  
\frac{1}{E_3 - \frac{{\cal A}+2}{2({\cal A}+1)} q^2 - \frac{{{\cal A}}+1}{2{{\cal A}}}k^2}
\right),\\
\label{tauaa}
&&\tau^{-1}_{AA}\left(E_3 - \frac{{\cal A}+2}{4{\cal A}}q^2\right) = 
\int d^D k \left(\frac{1}{- |E_2^{AA}|- k^2} - \frac{1}{E_3 - \frac{{\cal A}+2}{4{\cal A}} q^2 - k^2}\right),
\end{eqnarray}
\end{small}
where $E_2^{AB}$ and $E_2^{AA}$ are the two-body energies of the bound $AB$ and $AA$ systems, respectively. 

In order to solve the above equations we will need two ingredients from an article published by Danilov \cite{danilov} almost ten years before the seminal work of Efimov. The first idea is that {\it (i) the properties of Efimov states are determined by the region of very large momenta}. In such region all energy scales $E_3$, $E_2^{AA}$ and $E_2^{AB}$ are set to zero. In this situation, one can obtain closed forms for the amplitudes $\tau_{AB}$ and $\tau_{AA}$:
\begin{small}
\begin{eqnarray}
\nonumber
\tau^{-1}_{AB}\left(-\frac{{\cal A}+2}{2({\cal A}+1)}q^2\right) &=& 
- q^{D-2} \left( \frac{{\cal A}+2}{2({\cal A}+1)}  \right)^{{D/2-1}} 
\; \left(\frac{2{\cal A}}{{\cal A}+1}\right)^{{D}/{2}}\\
&\times& \Gamma \left(D/2-1\right)
\Gamma\left(2-D/2\right) \frac{\pi^{{D}/{2}}}{\Gamma({D}/{2})},
\end{eqnarray}
\begin{eqnarray}
\tau^{-1}_{AA}\left(- \frac{{\cal A}+2}{4{\cal A}}q^2\right) &=&  - q^{D-2} 
\left( \frac{{\cal A}+2}{4{\cal A}}  \right)^{{D/2-1}} 
\, 
\Gamma \left({D/2-1}\right) \Gamma\left({2-D/2}\right)\frac{\pi^{{D}/{2}}}{\Gamma\left({D}/{2}\right)},
\end{eqnarray}
\end{small}
where $\Gamma(z)$ is the gamma function, defined for all complex numbers $z$ except for the non-positive integers. It is important to note here that functions $\chi_A(q)$ and $\chi_B(q)$ are invariant under a scale transformation, which can be verified performing the following replacements $q\rightarrow\lambda q$ and $k\rightarrow\lambda k$ in Eqs. (\ref{chia}) and (\ref{chib}) - this is the second ingredient from  Danilov \cite{danilov} - {\it (ii) such scale invariant functions have a power law form that can be written as}
\begin{equation}
\chi_A(q) = C_A \, q^{r+is} \,\,\,{\rm and}\,\,\, \chi_B(q) = C_B \, q^{r+is},
\label{ansatz}
\end{equation}
where $C_A$, $C_B$, $r$ and $s$ are real numbers. These solutions are the well-known log-periodic functions, associated with the infinitely many three-body bound states in the Efimov limit. Thus, the  Efimov effect is related to the appearance of an imaginary $s$ factor which also is associated to the loss of conformal invariance mentioned in Ref. \cite{mohapatra}.

The replacement of Eq.~(\ref{ansatz}) in Eqs.~(\ref{chia}) and (\ref{chib}) leads to a complex homogeneous 
linear matrix equation for the coefficients $C_A$ and $C_B$. The parameters $r$ and $s$ 
are found by solving the corresponding characteristic equation. The real part of the 
exponent is given by the ansatz $r = 1 - D$ for all $D$, which removes any ultraviolet divergence.
For $D=3$ and ${{\cal A}}=1$ one obtains $s=1.00623$, value that agrees with the well 
known result from Efimov \cite{efimov}. Moreover, for $r=1-D$, the characteristic equation reads:
\begin{eqnarray} 
\nonumber
&&{\cal F}_D \left[{\cal A} \, I_1({\cal A},s)  + 2  
\left( \frac{4{\cal A}}{{\cal A}+2} \right)^{{D/2-1}} 
{\cal F}_D \, I_2({\cal A},s) I_3({\cal A},s) \right]\\
&&= \left( \frac{{\cal A}+2}{2({\cal A}+1)}  \right)^{{D/2-1}}  
\left(\frac{2{\cal A}}{{\cal A}+1}\right)^{{D}/{2}} ,
\label{eqfinal}
\end{eqnarray}
where
\begin{equation}
{\cal F}_D=\frac{1}{\Gamma\left({D/2-1}\right) 
\Gamma\left({2-D/2}\right)},
\end{equation} 
and 
\begin{eqnarray}
\hspace{-0.5cm}
I_{1}({\cal A},s) \!&=&\! \int^{\infty}_0 \!dz\frac{z^{is}}{z} 
\log \left[ \frac{(z^{2}+1)({\cal A}+1) +2z}{(z^{2}+1)({\cal A}+1) -2z}\right], \\
\hspace{-0.5cm}
I_{2}({\cal A},s) \!&=&\! \int^{\infty}_0 \!dz\frac{z^{is}}{z} 
\log \left[ \frac{2{\cal A}(z^{2}+z) + {\cal A}+1}{2{\cal A}(z^{2}-z)+{\cal A}+1}\right], \\
\hspace{-0.5cm}
I_{3}({\cal A},s) \!&=&\! \int^{\infty}_0 \!dz\frac{z^{is}}{z} \log \left[ \frac{2{\cal A}(1+z) 
+ ({\cal A}+1)z^{2}}{2{\cal A}(1-z) + ({\cal A}+1)z^{2}}\right] ,  
\end{eqnarray}
which are the same integrals found in Ref. \cite{yamashitapra2013} for the $D=3$ problem. We note
that the result $r = 1 - D$ is exact. 

\begin{figure}[htb!]
\centering
\includegraphics[width=10cm]{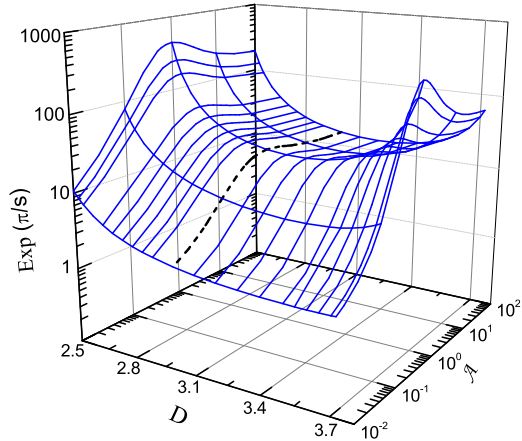}
\caption{Discrete scaling factor as a function of the mass ratio ${\cal A}=m_B/m_A$, and dimension $D$. The black dashed line shows the results for 3D. Figure originally published in Ref. \cite{derickpra}.} 
\label{efimovfactor}
\end{figure}

The solution of Eq. (\ref{eqfinal}) for a given mass ratio and $D$ returns the Efimov discrete scaling factor $s$, which determines the three-body energy and root-mean-square hyperradius ratios between two consecutive states given, respectively, by exp$[2\pi/s]$ and exp$[\pi/s]$ (represented in Fig. \ref{efimovfactor} as a function of ${\cal A}$ and $D$). 

\begin{figure}[htb!]
\includegraphics[width=9cm]{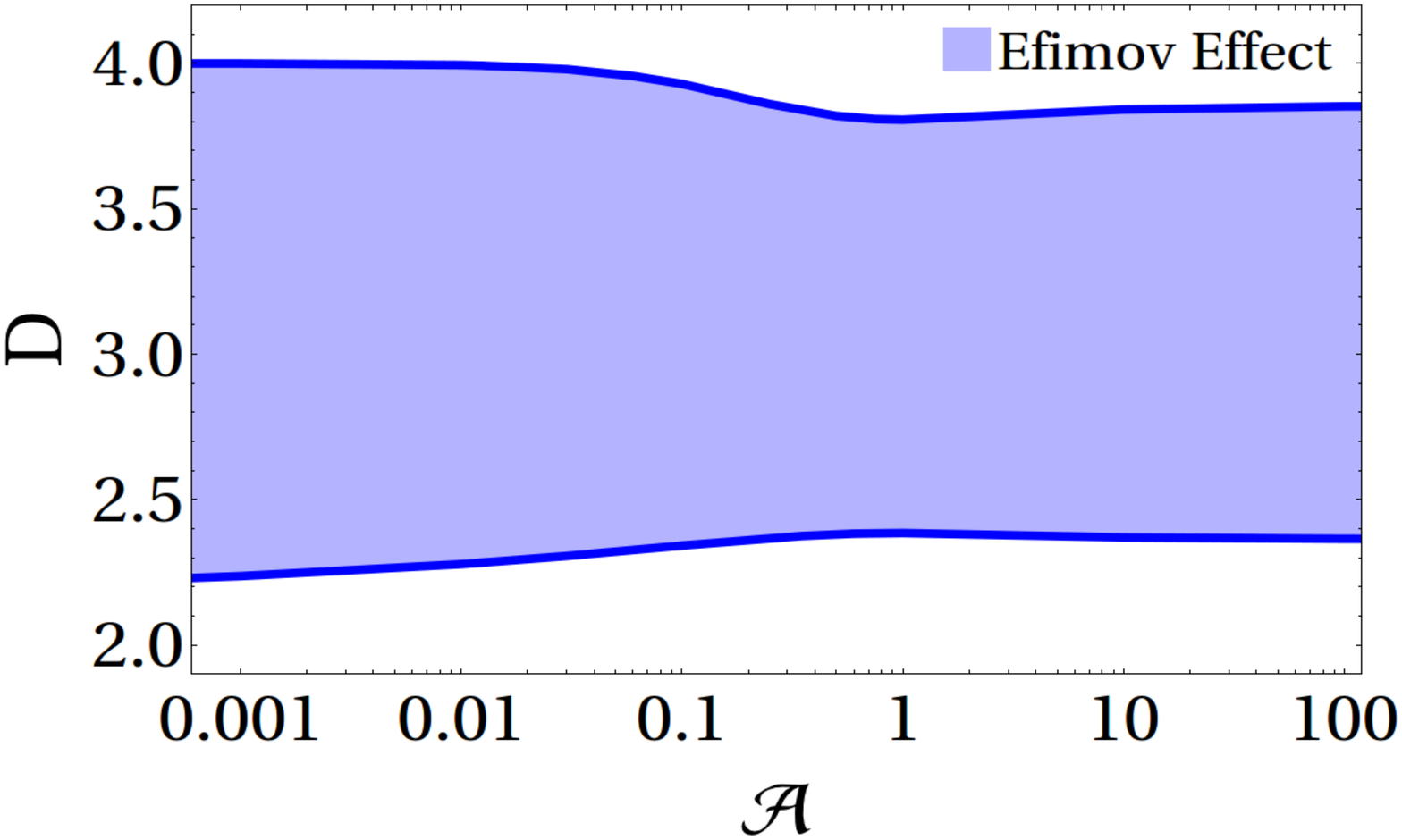}
\caption{Regions (in blue) where there is a real solution for the scaling factor, $s$, solution
to the Eq.~(\ref{eqfinal}); outside this ``dimensional band'', the Efimov effect does not exist. Figure originally published in Ref. \cite{derickpra}.} 
\label{boundaries}
\end{figure}

For a given ${\cal A}$ there is an interval for $D$ where there is a real solution for $s$ in Eq. (\ref{eqfinal}). Then, we can define an ``Efimov band'' - a dimensional interval as a function of ${\cal A}$ where the Efimov effect exists. This is plotted in Fig. \ref{boundaries}.

Note that an imaginary $s$ means the loss of scale invariance in Eqs.~(\ref{chia}) and (\ref{chib}). However, this does not give the physical reason for the existence of an Efimov state. This will be discussed in the next section.

\section{A less ideal case: three-body system in the Born-Oppenheimer (BO) approximation}

In the previous section, for a given ${\cal A}=m_B/m_A$ we showed a dimensional interval where Efimov states exists and calculated the respective discrete scaling factor, $s$. However, as pointed out, we did not discuss the physical reasons for the disappearance of such states. This discussion will be made in this section with a less ideal situation: the physical scales are now present, but there is a condition for ${\cal A}$. Part of the content of this section was published in \cite{derickjpb}. In the Landau and Lifshitz book on nonrelativistic quantum mechanics \cite{landau}, one can read: ``to reveal certain properties of quantum-mechanical motion it is useful to examine a case which, it is true, has no direct physical meaning: the motion of a particle in a field where the potential energy becomes infinite at some point (the origin) according to the law $U(r)\sim-\beta/r^2\,\,(\beta>0)$''. This book was first published in English language in 1958 and the transcribed sentence appears at the the beginning of the subsection ``fall of a particle to the centre'', in which the possible solutions of the Schr\"odinger equation for such a potential are studied. 

Contrary to the first part of the exert of Landau's book, where it is written that the potential proportional to $r^{-2}$ ``there is no physical meaning'', the `fall of a particle to the centre'' is the essence of the Efimov effect. In Ref. \cite{fonseca}, Fonseca, Redish and Shanley  used the Born-Oppenheimmer approximation to study a three-body system in 3D and derived explicitly the $1/\rho^2$ ($\rho$ is the hyperradius) effective potential responsible for the appearance of the Efimov effect. In this section we derive the $D$-dimensional effective potential responsible for the fall to the center. The BO approximation requires that ${\cal A}\ll1$.

The BO approximation allows to separate the full three-body Schr\"odinger equation into two equations, one for the heavy-light subsystem
\begin{small}
\begin{eqnarray}
\left[ -\frac{\hbar^{2}}{2\mu_{B,AA}}\nabla^{2}_{r} 
+ V_{AB}(\vec{r}-\frac{\mu_{AA}}{m_{A}}\vec{R})  
+ V_{AB}(\vec{r}+\frac{\mu_{AA}}{m_{A}}\vec{R})\right]\psi_R(\vec{r}) 
= \epsilon(R) \psi_R(\vec{r}), 
\label{light1}
\end{eqnarray}
\end{small}
and another for the two heavy particles:
\begin{equation}
\left[-\frac{\hbar^{2}}{2\mu_{AA}}\nabla^{2}_{R}  + V_{AA}(\vec{R}) + \epsilon(R) \right]\phi(\vec{R}) = 
E_3 \phi(\vec{R}),
\label{heavy}
\end{equation}
where $V_{AB}$ and $V_{AA}$ are the $AB$ and $AA$ two-body interactions, respectively and the reduced masses are given by $\mu_{B,AA}=2m_A m_B/\left(2m_A+m_B\right)$ and $\mu_{AA}=m_A/2$. As usual with the Born-Oppenheimer approximation, $R$ enters as a parameter in Eq.~(\ref{light1}) and can be used as a labelling index, and the eigenvalue $\epsilon(R)$ of the heavy-light equation enters as an effective potential in Eq.~(\ref{heavy}) for the heavy-heavy system. The solution of the light particle equation returns the effective potential, $\epsilon(R)$, responsible for the appearance of the Efimov effect.

We employ a contact interaction with strength $\lambda$ for the heavy-light potential $V_{AB}$. In order to solve such equation we will follow the recipe from Ref. \cite{bellottijpb2013} but replacing the calculations to a $D$-dimensional space. Thus, the light particle Eq.~(\ref{light1}) written in momentum space reads
\begin{equation}
\frac{q^2}{2\mu_{B,AA}} \tilde{\psi}_R(\vec{q}) 
+ \lambda \left[ e^{i \vec{q}\cdot\vec{R}/2} \, A_R^{(+)}
+ e^{-i\vec{q}\cdot\vec{R}/2} \, A_R^{(-)}\right] 
= \epsilon(R) \tilde{\psi}_R(\vec{q}),
\label{mom}
\end{equation}

where
\begin{equation}
A_R^{(\pm)}=\int \frac{d^Dq}{(2\pi)^D} \, 
e^{ \mp i \vec{q}\cdot\vec{R}/2 } \, \tilde{\psi}_R(\vec{q}),
\label{Apm_R}
\end{equation}
with $\tilde{\psi}_R(\vec{q})$ being the Fourier transform of $\psi_R(\vec{r})$, defined as
\begin{equation}
\psi_R(\vec{r}) = \int \frac{d^Dq}{(2\pi)^{D/2}} \, e^{i \vec{q}\cdot\vec{r}} \, \tilde{\psi}_R(\vec{q}) .
\end{equation}

The eigenvalue $\epsilon(R)$ can be determined as follows \cite{bellottijpb2013}. 
Initially, one rewrites Eq.~(\ref{mom}) as
\begin{equation}
\tilde{\psi}_R(\vec{q}) = \lambda \,  
\frac{e^{i \vec{q}\cdot\vec{R}/2}  A_R^{(+)}+ e^{-i\vec{q}\cdot\vec{R}/2} \, A_R^{(-)}}{\epsilon(R) - \frac{q^2}{2\mu_{B,AA}}} .
\end{equation}
Next, one eliminates $\tilde{\psi}_R(\vec{q})$ in favor of $A_R^{(\pm)}$ using Eq.~(\ref{Apm_R}):
\begin{equation}
A_R^{(\pm)} = \lambda\int \frac{d^Dq}{(2\pi)^D} \; \frac{A_R^{(\pm)}
+ e^{\mp i\vec{q}\cdot\vec{R}} \, A_R^{(\mp)}}{\epsilon(R) - \frac{q^2}{2\mu_{B,AA}}} .
\end{equation}
Then, it is easily shown that for nontrivial solutions,  $A_R^{(\pm)}\neq 0$, the eigenvalue
is given by the transcendental equation:
\begin{equation}
\label{trans}
\frac{1}{\lambda} = \int \frac{d^Dq}{(2\pi)^D}
\, \frac{\cos(\vec{q}\cdot\vec{R}) + 1 }{\epsilon(R)-\frac{q^2}{2\mu_{B,AA}}}.
\end{equation}
The integral is divergent but the divergence can be dealt with by eliminating the strength 
$\lambda$ in favor of the two-body binding energy. That is, assuming that the two-body subsystem
contains a bound state with energy $E_{AB}\equiv-|E_2|$
\begin{equation}
\label{bound}
\frac{1}{\lambda} = - \int \frac{d^Dq}{(2\pi)^D}
\, \frac{ 1 }{|E_2| + \frac{q^2}{2\mu_{B,AA}}},
\end{equation}
and using this to replace $1/\lambda$ in Eq. (\ref{trans}) one obtains
\begin{equation}
\int \frac{d^Dq}{(2\pi)^D} \left[
\frac{\cos(\vec{q}\cdot\vec{R})+1}{\epsilon(R)-\frac{q^2}{2\mu_{B,AA}}}
+\frac{1}{|E_2|+\frac{q^2}{2\mu_{B,AA}}} \right] = 0.
\label{effpot}
\end{equation}

The integral in Eq. (\ref{effpot}) can be solved analytically; the
result is
\begin{eqnarray}
2^{D/2}{\bar R}^{1-D/2}|\bar{\epsilon}({\bar R})|^{(D-2)/4}
K(D/2-1,{\bar R}|\bar{\epsilon}({\bar R})|^{1/2})
&& \nonumber \\ 
-  \pi\csc\left(D\pi/2\right) 
\frac{1}{\Gamma(D/2)} \,
\left(1-|\bar{\epsilon}
({\bar R})|^{D/2-1}\right) = 0&&, \;\;\;\;\;
\label{fullequation}
\end{eqnarray}
where $|\bar{\epsilon}(R)| = |\epsilon(R)|/|E_{2}|$ and ${\bar R}=R/a$ 
with $a=\sqrt{\hbar^{2}/{2\mu_{B,AA}|E_{2}|}}$. $K(z)$ is the modified Bessel function of the second kind.

The effective potential $\bar{\epsilon}(R)$ is obtained from 
the transcendental equation in Eq.~(\ref{fullequation}) that can be solved for 
a given mass ratio and dimension $D$. The small distance regime of the effective potential provides the condition for the appearance of the Efimov effect. 

Performing the limit ${\bar R}\rightarrow0$, the effective potential for values of $D$ in the interval $2<D<4$, is given by
\begin{equation}
|\bar{\epsilon}_{R \rightarrow 0}(\bar{R})| = \frac{g(D)}{\bar{R}^{2}} ,
\label{gamma}
 \end{equation}
where $g(D)$ is the solution of the transcendental equation
\begin{equation}
g(D) = \left[-\frac{ \pi \csc(D \pi/2)}
{ 2^{D/2} \Gamma({D}/{2}) K(D/2-1,\sqrt{g(D)}) } \right]^{\frac{4}{2-D}} .
\end{equation}
For $D=3$, one obtains
\begin{equation}
g(3) = 0.3216,
\end{equation}
which reproduces the $D=3$ results \cite{fonseca,mathias}. Solely the form $1/\bar{R}^{2}$ is not sufficient for the appearance of the Efimov effect. The numerator, given by $g(D)$, also plays a central role here.

\begin{figure}[htb!]
\centering
\includegraphics[width=9cm]{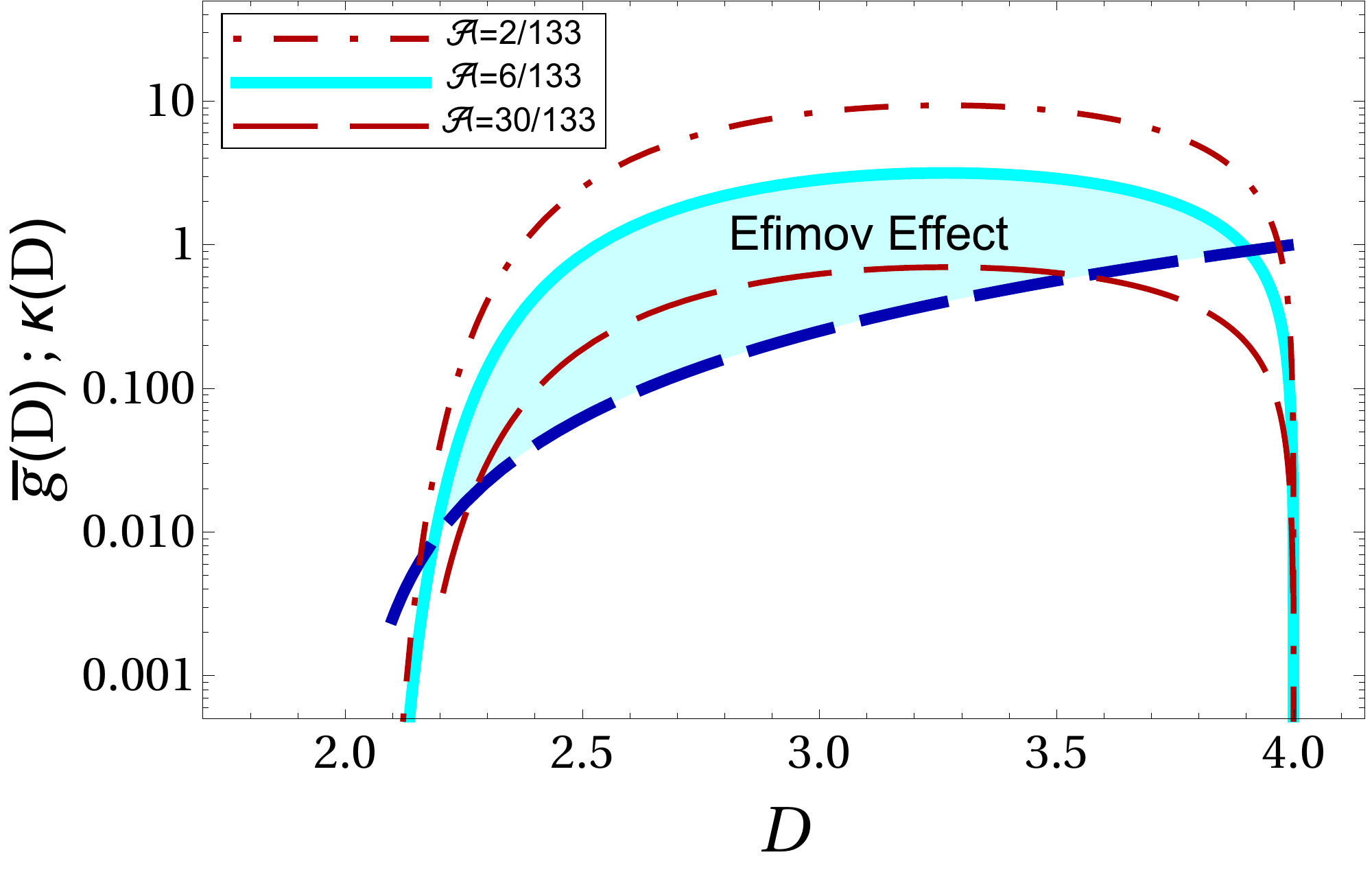}
\caption{The shadowed area shows the region where the Efimov effect is allowed according to 
the effective strength given by Eq. (\ref{gamma}). The blue dashed line is the critical 
strength given by $\kappa(D)=(D-2)^{2}/4$ Eq. (\ref{scalingD}), and the effective strength 
${\bar g}(D)=({\cal A}+2)g(D)/4{\cal A}$ is presented for three 
different mass ratios. The dimensional 
limit is $2.21<D<3.90$ for ${\cal A}=6/133$. Figure originally published in Ref. \cite{derickjpb}.} 
\label{figDlimits}
\end{figure}

The $D$-spherical Schr\"odinger equation \cite{martins} for zero total angular momentum is given by
\begin{equation}
\left[-\frac{\hbar^{2}}{2\mu_{AA}}\nabla^{2}_{R}  + \epsilon(R) \right]\phi(\textbf{R}) = E \phi(\textbf{R}),
\end{equation}
where the radial part of the Laplacian reads
\begin{equation}
\nabla^{2}_{R}  = \frac{d^{2}}{dR^{2}}+\frac{D-1}{R}\frac{d}{dR}.
\end{equation}
For two identical heavy particles and considering an infinitely high excited three-body state 
with energy $E_{3}\approx0$ we can write
\begin{equation}
\left[\frac{d^{2}}{dR^{2}}+\frac{D-1}{R}\frac{d}{dR}  - \frac{2\mu_{AA}}{\hbar^{2}} \epsilon(R) \right]\phi(R) = 0.
\end{equation}
The effective potential in the regions where the Efimov effect appears (small distances) is given by 
Eq. (\ref{gamma}), $-{\cal G}/R^{2}$, where ${\cal G}$ depends on the mass ratio and dimension. Replacing 
the asymptotic effective potential and using the ansatz $\phi(R)=R^{\delta}$ we can calculate the 
Efimov discrete scaling factor $s$ for a general $D$ $(2<D<4)$.
\begin{equation}
\delta= \frac{1}{2}\pm i \sqrt{\frac{{\cal A}+2}{4 {\cal A}}g(D) - (D/2-1)^{2}}=\frac{1}{2}\pm i s.
\label{scalingD}
\end{equation}
In order to have the Efimov effect, $\phi(R)$ should oscillate, thus $s$ should be real. 
As an example, Figure \ref{figDlimits} shows for a $^{133}$Cs-$^{133}$Cs-$^6$Li system, ${\cal A}=6/133$, 
the region where the Efimov effect is allowed. The difference of the $D$ limits of the BO approximation 
with the exact result \cite{derickjpb} is less than 2\%. Note that the effect of a finite $E^{D}_2$ is washed 
out here as the critical strength is obtained in the limit of ${\bar R}\rightarrow0$.

\section{General system: scales with compactified dimensions}

In this section we will consider a general three-body $AAB$ system, where physical scales are present and there is no restriction to $m_A$ and $m_B$. The transition from 3D to 2D is performed in momentum space and the energy spectrum comes from the solutions of STM-like integral equations derived from the Faddeev equations \cite{faddeev}. Part of the following content was first published in Refs. \cite{john,yamashitapra2013}. The dimensional transition is made as follows. Consider a generic three-body system described by relative Jacobi momenta $(p_x,p_y,p_z)$ with periodic boundary conditions along one direction (chosen to be the $z$-axis). Then, the relative momenta along the plane are given by $\vec p_\perp=(p_x,p_y)$ and
\begin{eqnarray}
p_z=\frac{n}{R_z} \ , \label{compact1}
\end{eqnarray}
with $n=0,\pm1,\pm2,\dots$. 

The length of the squeezed dimension corresponds to a radius, $R_z$, that interpolates between the 2D limit for $R_z\to 0$ and the 3D limit for $R_z\to \infty$. The 2D limit is achieved by increasing the gap between the momenta in the $z$-direction in such a way we cease the propagation along this direction. The choice of a periodic dimension is not essential as we may map the physics of other types of external confinement onto the system with periodic boundary conditions. We can use exactly the same idea with $p_x$ or $p_y$ to move continuously from 2D to 1D. In each case, the respective integral is then replaced by a sum in the integral equations.

The method briefly described above was first used in Ref. \cite{yamashitapra2013}. It presents a numerical difficulty when approaching the limit of large $R$'s. In this limit the number of terms in the sum should be increased, which affects considerably the total time of the numerical calculation. In order to circumvent this problem we inserted an angular decomposition of the kernel of the integral equations. Between the 3D and 2D limits, the decomposition inserts Legendre polynomials, which can be used to reduce the number of terms in the sum close to the large $R$'s. This improvement of our first version of the compactification method \cite{yamashitapra2013} can be found in appendix B of Ref. \cite{john}.

After quantizing the relative momenta $\vec q$ and $\vec p$ ($\vec q$ has its origin is the center-of-mass of a given pair and point towards the remaining particle and $\vec p$ connects the pair) in the $z$ direction and performing the angular decomposition, the coupled and subtracted integral equations for the spectator functions, $\chi_B$ and $\chi_A$, reads:
\begin{small}
\begin{eqnarray}
\label{zre5a}
&&\chi_B(\tilde q)=-2\,\tau_{AA;R_z}\left(E_3-\frac{{\cal A}+2}{4{\cal A}} \tilde q^2\right)\sum_m\int\frac{d^2p_\perp}{R_z} 
{\cal H}^{(1)}(\tilde q,\tilde p;E_3,-\mu^2) \chi_A(\tilde p)\\ \nonumber
&&\chi_A(\tilde q)=-\tau_{AB;R_z}\left(E_3-\frac{{\cal A}+2}{2({\cal A}+1)}  \tilde q^2 \right)\sum_m\int \frac{d^2p_\perp}{R_z} \left\{
{\cal H}^{(1)}(\tilde p,\tilde q;E_3,-\mu^2) \chi_B(\tilde p)\right.\\
&&\left.+ {\cal H}^{(2)}(\tilde q,\tilde p;E_3,-\mu^2) \chi_A(\tilde p)\right\}
\label{zre5b}
\end{eqnarray}
\end{small}
where  $\tilde q\equiv (\vec q_\perp,n)$, $\tilde p\equiv (\vec p_\perp,m)$ and
\begin{equation}
\nonumber
\tilde q^2=q_\perp^2+{n^2\over R_z^2}\,\, ,\,\, \tilde p^2=p_\perp^2+{m^2\over R_z^2} \,\, {\rm and} \,\,  
\tilde q\cdot \tilde p=\vec q_\perp\cdot \vec p_\perp +\frac{n\;m}{R_z^2}
\end{equation}
The functions ${\cal H}^{(1)}(\xi,\eta;E_3,-\mu^2)$ and ${\cal H}^{(2)}(\xi,\eta;E_3,-\mu^2)$ are given by
\begin{eqnarray}
&&{\cal H}^{(1)}(\xi,\eta;E_3,-\mu^2)=G^{(1)}_{0R_z}(\xi,\eta;E_3)-G^{(1)}_{0R_z}(\xi,\eta;-\mu^2)\\
&&{\cal H}^{(2)}(\xi,\eta;E_3,-\mu^2)=G^{(2)}_{0R_z}(\xi,\eta;E_3)-G^{(2)}_{0R_z}(\xi,\eta;-\mu^2),
\end{eqnarray}
where the resolvents are defined by:
\begin{eqnarray}
&&\left[G^{(1)}_{0R_z}(\tilde q,\tilde p;E)\right]^{-1}=
E-\tilde p^2-\tilde q\cdot \tilde p-\frac{{\cal A}+1}{2{\cal A}}\tilde q^2 
\\
&&\left[G^{(2)}_{0R_z}(\tilde q,\tilde p;E)\right]^{-1}= E
-\frac{\tilde q\cdot \tilde p}{{\cal A}}-\frac{{\cal A}+1}{2{\cal A}}(\tilde q^2+\tilde p^2)
\label{greenq2d}
\end{eqnarray}
The two-body amplitudes for finite $R_z$ are given by
\begin{small}
\begin{equation}
R_z \,\tau^{-1}_{A\beta;R_z}(E)
 =2\,m_{A\beta}\,\left\{\sum_n\int  
\frac{d^2p_\perp}{ \tilde E-
p_\perp^2-\frac{n^2}{R_z^2}}
-\sum\int  
\frac{d^2p_\perp}{\tilde E_{A\beta}-
p_\perp^2-\frac{n^2}{R_z^2}} \right\}
\ , \label{rtauab}
\end{equation}
\end{small}
with $\beta\equiv A$ or $ B$, $\tilde E=2\,m_{A\beta} E$ ($E<0$) and $\tilde E_{A\beta}=2\,m_{A\beta} E_{A\beta}$ 
and we chose the bound-state pole at $E_{A\beta}$ for each $R_z$. The reduced mass is $m_{A\beta}=m_A\, m_\beta/(m_A+ m_\beta)$.
Performing the analytical integration over $\vec p_\perp$ and performing the sum, we get that
\begin{eqnarray}
\tau_{A\beta;R_z}(E)=R_z\left[
4\pi\,m_{A\beta}\ln\left({\sinh\pi\sqrt{-2\,m_{A\beta}E}\,R_z\over\sinh\pi\sqrt{-2\,m_{A\beta}E_{A\beta}}\,R_z}\right)\right]^{-1}
\label{tauq3d}\end{eqnarray}
In  the limit of $R_z\to\infty$  the two-body amplitudes for $AA$ and $AB$ reduces to the known 3D expressions.

\begin{figure}[htb!]
\centering
\includegraphics[width=9cm]{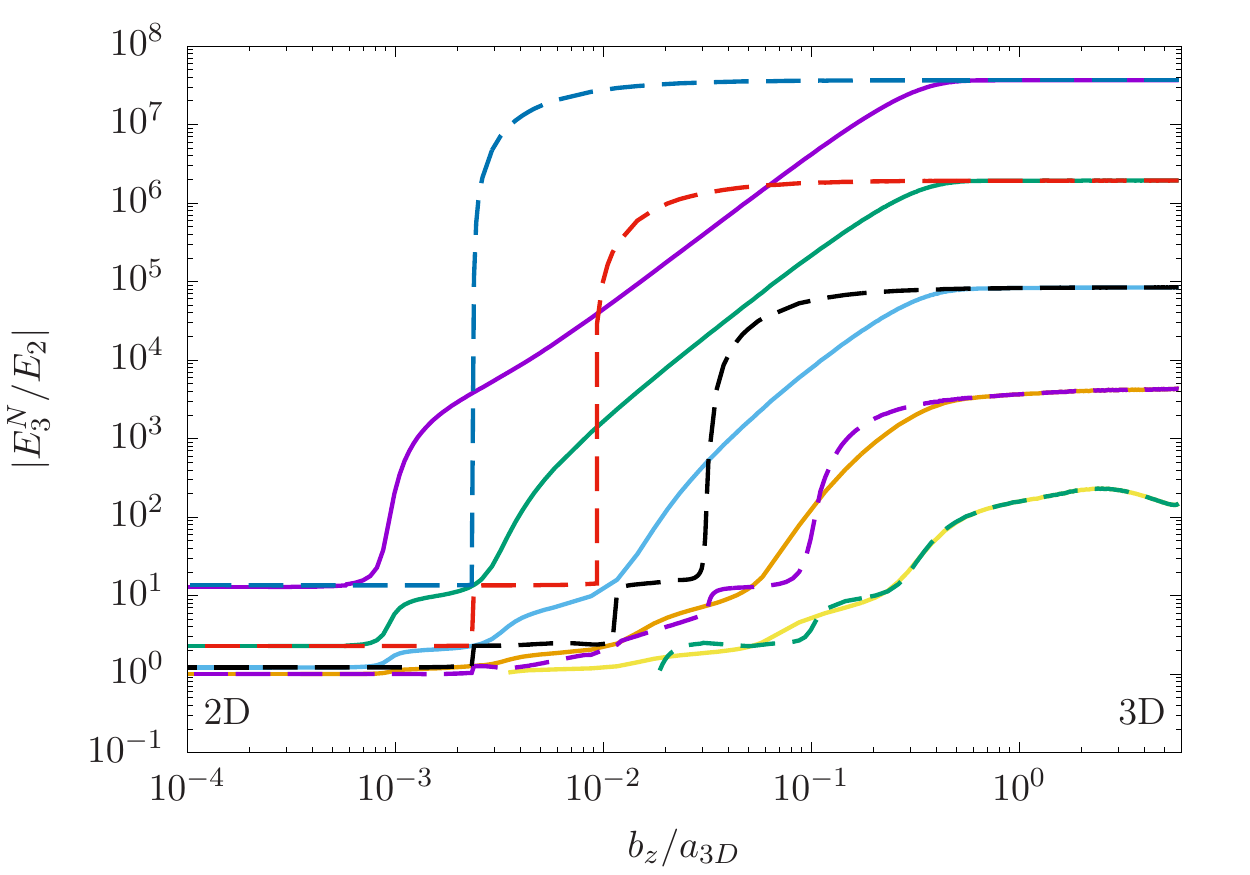}
\caption{Trimer energies plotted in units of the two-body energy for $m_B/m_A = 6/133$ as functions of $b_{z}/a_\textrm{3D}$. For the solid lines the two-body energy varies with $b_z$ while for the dashed lines it is kept constant. Figure originally published in Ref. \cite {john}.} 
\label{3spectrum}
\end{figure}

The solution of Eqs. (\ref{zre5a}) and (\ref{zre5b}) returns the three-body energy spectrum. The transition from 3D to 2D for $m_B/m_A = 6/133$ is showed in Fig. \ref{3spectrum}. The compactification parameter, $R_z$, was related with the oscillator length, $b_z=\sqrt{\hbar/\mu_{AB}\omega_z}$ ($\mu=m_Am_B/(m_A+m_B)$ and $\omega_z$ is the frequency of the oscillator) in such a way the three-body energies are given as a function of $b_{z}/a_\textrm{3D}$ ($E_2=\hbar^2/\mu_{AB}a_\textrm{3D}^2$). In the considered situation the subsystem $AA$ is not interacting.

\section{Conclusion}

In this article, we showed how the Efimov effect is affected by a continuous transition of the spatial dimensions. We started with an ideal situation, where the study of an infinite high excited state provides an easy way to find the dimensional limits for the existence of Efimov states, as well as the discrete Efimov scaling factor. Then the BO approximation was used to extract the effective potential that generates the fall to the center. In the last case, we considered a general situation where we introduced our compactification method.

We still need much more studies to understand what is the physical meaning of a fractional $D$ in Efimov states. Besides we have already some estimates on how the effective dimension changes according to the oscillator parameter \cite{derickpra,christensen} a direct relation of $D$ with some trap parameter should still be found.

From the experimental point-of-view, several techniques to construct 3D, 2D or 1D ultracold atomic traps are known long ago. The number of Efimov excited states, as well as the ratio between consecutive energy levels, may be controlled by the mass ratio of the atoms and many experiments have already been made considering heteronuclear species \cite{naidon,mathias,ulmanis}. The reduced dimension starts to be felt by the Efimov trimer once its two-body scattering length becomes comparable to the size of the trap, e.g., the oscillator length of a harmonic trap. In this situation, the degrees of freedom in one or two directions are suppressed and the system behaves as being in 2D or 1D. Once the Efimov effect does not exist in 2D and 1D, the disappearance of the 3D Efimov excited states can be observed by measuring the peaks of the three-body recombination loss. Alternatively, it is also possible to measure directly the energy spectrum \cite{dornertrimer,dornerdimer,jochim}. 

Finally, as we showed in this article, dimensional effects in few-body systems are a very interesting topic to be explored by the few-body community. Many interesting aspect of few-body physics in deformed geometries may already be studied by experimentalists using the currently available technology. 

\section{Acknowledgements}

This work was partly supported by funds provided by the Brazilian agencies Funda\c{c}\~{a}o de Amparo \`{a} Pesquisa do Estado de S\~{a}o Paulo - FAPESP grants no. 2016/01816-2, Conselho Nacional de Desenvolvimento Cient\'{i}fico e Tecnol\'{o}gico - CNPq grant no. 302075/2016-0(MTY), Coordena\c{c}\~{a}o de Aperfei\c{c}oamento de Pessoal de N\'{i}vel Superior - CAPES grant no. 88881.030363/2013-01. I would like to especially thank my collaborators D.S. Rosa, J.H. Sandoval, T. Frederico and G. Krein for all discussions during the development of the articles reported in this review.

\end{document}